\documentclass[a4paper,12pt]{article}
\usepackage{latexsym}
\usepackage{amsmath}
\usepackage{amsfonts}
\usepackage{amssymb}
\usepackage{graphicx}
\usepackage{color}
\usepackage[LGR,T1]{fontenc}
\usepackage[greek,frenchb,english]{babel}

\title {On Jump-Diffusive Driving  Noise Sources \\ Some Explicit Results and Applications}
\date{}

\begin{document}
 \author{ M.-O. Hongler$^{1}$\thanks{Corresp. author: max.hongler@epfl.ch} \,\,  \& \,\, R. Filliger$^{2}$\\
{\small $^{1}$LPM, Ecole Polytechnique Fédérale de Lausanne, 1015 Lausanne, Switzerland, }\\
{\small $^{2}$IEM, Bern University of Applied Sciences, 2501 Biel, Switzerland.}}

\maketitle

\abstract{

\noindent  We study some linear and nonlinear shot noise models where the jumps are drawn from a compound Poisson process with jump sizes following an Erlang-$m$ distribution. We show that the associated Master equation can be written as a spatial $m^{{\rm th}}$ order partial differential equation without integral term. This differential form is valid for state-dependent Poisson rates and we use it to characterize, via a mean-field approach, the collective dynamics of a large population of pure jump processes interacting via their Poisson rates. We explicitly show that for an appropriate class of interactions, the speed of a tight collective traveling wave behavior can be triggered by the jump size parameter $m$. As a second application we consider an exceptional class of stochastic differential equations with nonlinear drift, Poisson shot noise and an additional White Gaussian Noise term, for which explicit solutions to the associated Master equation are derived.
}

\vspace{0.4cm}
 Compound Poisson noise sources with Erlang jump distributions. Higher order partial differential equations. Lumpability of Markov processes. Mean-field approach to homogeneous multi-agents systems. Flocking behavior of multi-agents swarms.

\vspace{0.1cm}
\noindent {\bf Mathematics classification numbers}.

\begin{itemize}
\item[]  {\bf 60G20}  {\it Generalized stochastic processes}
\item[]  {\bf 60H10} {\it Stochastic ordinary differential equations }
\item[]  {\bf 82C31}   {\it Stochastic methods (Fokker-Planck, Langevin, etc.) }
\item[]  {\bf 60K35} {\it Interacting random processes; statistical mechanics type models}

    \end{itemize}

\section{Introduction}

\noindent On the real line  $\mathbb{R}$,  we shall  consider scalar  time-dependent Markovian stochastic processes $X_t$, ($t \in \mathbb{R}^{+}$ is the time parameter) characterized by  stochastic differential equations (SDE) of the form:

\begin{equation}
\label{SDE1}
\left\{
\begin{array}{l}
dX_t = - f(X_t)  dt + \sigma(X_t,t)dW_t + q_{X_t, t}, \\ \\
X_0 = x_0,
\end{array}
\right.
\end{equation}

\noindent where $W_t$ is a standard Wiener process with diffusion coefficient $\sigma(x,t)$, $q_{X_t,t}$ stands for a compound Poisson process (CPP) with Poisson rate $\lambda(X_t,t)$ and jump sizes drawn from a given probability density $\varphi(x)$ and where the drift $-f(x)$ reflects the deterministic behavior of the system. If necessary (i.e., if $\sigma$ is space dependent), we will interpret \eqref{SDE1} in the Itô sense. Accordingly, the Master equation governing the evolution of the conditional probability density function (pdf) $P(x,t|x_0,0) =  {\rm Prob} \left\{X(t) \in \left[x, x+ dx\right]|x_0,0\right\}$ reads \cite{COX}:

\begin{equation}
\label{MARKOGEN}
\begin{array}{l}
\partial_t P(x,t |x_0,0) = \partial_{x} \left[ f(x) P(x,t |x_0,0) \right] + {1 \over 2} \partial_{xx} \left[\sigma^{2}(x,t)   P(x,t |x_0,0)\right] \\ \\
\qquad \quad - \lambda(x,t) P(x,t |x_0,0)  +   \int_{-\infty}^{\infty} \varphi(x-z) \lambda(z,t) P(z, t |x_0,0) dz.
\end{array}
\end{equation}

\noindent  Note that for $\lambda(x,t)\equiv 0$, the solution to Eq.(\ref{SDE1}) is a diffusion process with continuous trajectories. In the generic case where the Poisson rates are strictly positif, these trajectories show jumps and hence are discontinuous.

\noindent Due to its  extremely  wide range of potential applications,  Eq.(\ref{SDE1}) together with  Eq.(\ref{MARKOGEN})   deserved a long and still growing list of research records. In the last decade,  quite a few new  contributions became available (a non exhaustive list is \cite{DENISOV,Denisov2009,ELIAZAR,DALEY,DALEY06,BALAZS,PERRY}).  The goals were either to write classes of explicit expressions for means, variances, Laplace transforms or even for $P(x,t |x_0,0)$ or to express conditions ensuring the existence of finite time-invariant (i.e. stationary) probability measures. Our goal here is to add some new information to this general effort by:

\begin{itemize}
  \item[a)]  Deriving a new higher order partial differential equation -- equivalent to \eqref{MARKOGEN} --  valid when the jumps of the CPP are drawn from an Erlang-$m$ probability law:
\begin{equation}
\label{ERLANG}
\varphi(x) =  {\cal E}(m, \gamma; x):= {\gamma ^{m} x^{m-1} e^{- \gamma x}\over \Gamma (m) } \mathbb{\chi}_{x \geq 0}, \qquad m=1,2,\cdots,
\end{equation}

\noindent with rate parameter $\gamma>0$ and where $\mathbb{\chi}_{x \geq 0}$ is the indicator function of the event $\{x \geq 0\}$.
  \item[]
  \item[b)]  Constructing a new soluble class of multi-agents dynamics in which  agents with pure jumps (i.e. $\sigma(x,t) \equiv 0$)  interact via their inhomogeneous Poisson rates $\lambda(x,t)$  and where the jumps are drawn from $\varphi(x)$  taken as an  Erlang-$2$ distribution.
  \item[]
   \item[c)]  Solving explicitly  Eq.(\ref{MARKOGEN})  when  $f(x) = \beta \tanh(\beta x)$,  $\sigma =1$, and the jump sizes are symmetric: $\varphi(x) = \varphi(-x)$.
   \item[]
\end{itemize}

\section*{Pure jump processes with Erlangian jump sizes}

\noindent Consider the dynamics in \eqref{SDE1} with inhomogeneous Poisson rates $\lambda(x,t)$ and Erlangian jumps distribution with parameter $m$ as defined in Eq.(\ref{ERLANG}). In this case, the governing Master equation for $P_m(x,t)=P_m(x,t |x_0,0)$ reads:
{\small
\begin{eqnarray}
\label{ERLANGm}
\partial_t (P_m(x,t)) - \partial_x\big[f(x) P_m(x,t)\big] - {1 \over 2} \partial_{xx} \left[\sigma^{2}(x,t)   P_m(x,t)\right]& = &
\nonumber \\
&&
\hspace*{-105mm}- \lambda(x,t) P_m(x,t) + \int_{-\infty}^{x} {\gamma ^{m} (x-z)^{m-1} e^{- \gamma (x-z)}\over \Gamma (m)}  \lambda(z,t) P_m(z, t) dz.
\end{eqnarray}}

\vspace{0.2cm}
\noindent {\bf Proposition 1}

\noindent For sufficiently smooth deterministic drift $f(x)$, Poisson rates $\lambda(x,t)$ and diffusion coefficient $\sigma^2(x,t)$ (all at least $m$ times differentiable with respect to $x$), the integral form of the Master equation \eqref{ERLANGm} can be rewritten as  the $m^{{\rm th}}$-order spatial differential equation\footnote{We suppress the arguments $x$ and $t$ in $f(x)$, $\lambda(x,t)$, $\sigma(x,t)$ and $P_m(x,t |x_0,0)$. }:
%
%
%
\begin{equation}
\label{DIFFOSALT2}
\left[ \partial_x + \gamma \right]^{m} \Big( \partial_t P_m - \partial_x \left[f\cdot P_m\right] - \frac{1}{2} \partial_{xx} \left[\sigma^{2}   P_m\right] \Big) =
\big[\gamma^{m}-\left[ \partial_x + \gamma \right]^{m}\big]  \big(\lambda \cdot P_m \big)
\end{equation}
Moreover for $\lambda(x,t)=\lambda(x)$ and $\sigma^2(x,t)=\sigma^2(x)$ a stationary distribution to \eqref{ERLANGm} necessarily verifies:
\begin{equation}
\label{DIFFOSALT3}
-\left[ \partial_x + \gamma \right]^{m} \Big( \partial_x \left[f\cdot P_m \right]+ \frac{1}{2}  \partial_{xx} \left[\sigma^{2}   P_m\right] \Big) =
\big[\gamma^{m}-\left[ \partial_x + \gamma \right]^{m}\big]  \big(\lambda \cdot P_m \big).
\end{equation}

\vspace{0.2cm}\noindent  The proof of Proposition 1 is given in Appendix A. For arbitrary drift terms and Poisson rates, explicit solutions to Eq.(\ref{DIFFOSALT2}) or Eq.(\ref{DIFFOSALT3})  are obviously difficult to derive. For convenience and later use,  let us  briefly list a few situations with $\sigma^2=0$ yielding  tractable solutions.


\subsubsection*{Stationary solutions}

\noindent Here we suppose that the large $t$ limit of $P_m$ exists and we write $P_{s,m}(x)=\lim_{t\rightarrow \infty}P_{m}(x,t)$ for normalizable solutions to \eqref{DIFFOSALT3}.

\noindent $\bullet$ For  $m=1$, $\lambda = \lambda(x)$,  $\sigma(x,t)=0$ and drift force $f(x)$, we have by \eqref{DIFFOSALT3}:
\begin{equation}
\left[\partial_x + \gamma \right] \big( \partial_x [f(x)P_{s,1}] \big)=\partial_x ( \lambda(x) P_{s,1})
\end{equation}
with the well known solution
\begin{equation}
P_{s,1}(x) ={ {\cal N} \over f(x)}
e^{- \gamma x +  \int^{x} { \lambda(\xi)\over f(\xi)} d\xi},
\end{equation}
\noindent and where ${\cal N}$ is the normalization factor. Clearly, the stationary regime $P_{1,s}(x)$  will actually be reached only when ${\cal N}< \infty$.

\vspace{0.5cm}
\noindent $\bullet$ For  $m=2$ we have
\begin{equation}
\label{DISER2}
\big[\partial_x + \gamma\big]^2 \big\{ \partial_x (f(x)P_{s,2})  \big\}=\big[\partial_x^2+2\gamma \partial_x\big] (\lambda(x) P_{s,2})
\end{equation}

\vspace{0.5cm}
\noindent Introducing the notation $P_{s,2}(x)= e^{- \gamma x} Q(x)\stackrel{not.}{=}e^{- \gamma x} Q$,  Eq.(\ref{DISER2}) takes, after elementary manipulations, the form:
%
%
{\begin{equation}
\label{MMQR}
f(x)[Q]_{xx} + 2[f(x)]_x[Q]_x+ [f(x)]_{xx}Q= \big[\lambda (x)Q\big]_x+  \lambda(x) \gamma Q.
\end{equation}}
\noindent Eq.(\ref{MMQR}) cannot be solved for general drift $f(x)$ and Poisson rate $\lambda(x)$. However, in the linear (Ornstein-Uhlenbeck) case with $f(x) = \alpha x$ and for constant rate $\lambda(x)=\lambda$, Eq.(\ref{MMQR}) reduces to:
\begin{equation}
\label{BESSEL}
\alpha x [Q(x)]_{xx} + \big(2 \alpha - \lambda \big) [Q(x)]_x  - \lambda \gamma Q(x)=0.
\end{equation}

\noindent Invoking \cite{ABRAMOWITZ}\footnote{See the entry 9.1.53 with $q=1/2$, $p= (\lambda/ \alpha -1)$, $p^{2} - \nu^{2} q^{2}=0$ and imaginary $\lambda$ and simplify once by $z$.}, the normalized stationary density  reads:
%
%
\begin{equation}
\label{SOLU}
P_{s,2} (x) =  \gamma
e^{{\lambda \over \alpha} - \gamma x} \left[ \alpha \gamma x \over \lambda \right]^{{{\lambda \over \alpha} - 1 \over 2}} \mathbb{I}_{{\lambda \over \alpha }-1} \left( 2 \sqrt{{\gamma \lambda \over \alpha}x } \right)
\end{equation}

\noindent where $\mathbb{I}_{\nu}(x)$ stands for the modified Bessel function of the first kind. Let us emphasize that Eq.(\ref{SOLU}) was also obtained in \cite{DALEY} by using Laplace transformations.

\vspace{0.3cm}
\noindent  $\bullet$ For general $m$, arbitrary drift $f(x)$ and constant Poisson rate $\lambda$, the resulting dynamics is  known as the  nonlinear shot noise process and has been  discussed e.g. in \cite{ELIAZAR}. In most cases, only the Laplace transform of $P_{s,m}(x)$ (resp. $P_{m}(x,t)$) can be given explicitly and, provided $P_{s,m}(x)$ exists,  the $j^{{\rm th}}$-order cumulant  $\kappa^{(j)}_{s,m}$ of $P_{s,m}$ can be calculated using the relations:
\begin{equation}
\label{REQ}
\left\{
\begin{array}{l}
\kappa^{(j)}_{s,m} = \int_{0}^{\infty} x^{j} \lambda {\Gamma(m, \gamma ;x)\over f(x)} dx, \qquad j=1,2,\cdots\\ \\
\Gamma(m,\gamma; x) := \int_{x}^{\infty}   {\cal E}(m, \gamma ;\xi)d\xi, \end{array}
\right.
\end{equation}
\noindent where $ \Gamma(m,\gamma; x) $ is the incomplete gamma function \cite{ELIAZAR}.


\subsubsection*{Time dependent solutions}

\noindent Time dependent solutions to \eqref{DIFFOSALT2} are available only for a restricted choice of drift terms $f$ and Poisson rates $\lambda$. Explicit  transient dynamics can be derived for constant drift $f(x) = k$, linear drift $f(x) = \alpha x$ and -- rather remarkably -- a non-linear interpolation between the two situations (discussed in section 3). The case of constant drift has been discussed in detail in \cite{Takacs}. The case for linear drift $f(x) = \alpha x$ and constant $\lambda$ is presented in \cite{DALEY}. Let us recall that in this latter case, the Laplace transform $\hat{P}_m (u,t): = \int_{\mathbb{R}^{+} } e^{- u x}P_m(x,t|x_0,0) dx$ read as:
   \begin{equation}
\label{LAPLIN}
\hat{P}_m (u,t)=\exp \left\{x_0 u e^{-\alpha t} - \lambda \int_{0}^{t}  \left(1 -  \left[{\gamma \over \gamma + \theta e^{ - \alpha (t-x)}}\right]^{m}\right) dx  \right\}.
\end{equation}
which can be inverted for $m=1$ yielding \cite{DALEY, PERRY}:
\begin{equation}
\label{LINM1}
\begin{array}{l}
P_1(x,t) = \chi_ze^{- \lambda t}\left\{ \delta(z)  + {\lambda \gamma \over \alpha}\left(e^{\alpha t} -1 \right)e^{- \gamma z} \, _1F_1\left( 1- {\lambda \over \alpha}, 2 \gamma \left[1- e^{\alpha t}  z\right] \right)\right\},
\end{array}
\end{equation}

 \noindent with $z = x - x_0 e^{-\alpha t}$, and where $\chi_z$ is the indicator function. Note that when $(1-\lambda/\alpha)=n$ is integer valued, $P_1(x,t)$ is an elementary function. Indeed, in this case $\,_1F_1(-n;b;z) $ reduces to the  $n^{{\rm th}}$-order generalized Laguerre polynomial  $L^{(1)}_{n}(z)$.

\section{Multi-agents systems and flocking}

As stated in the introduction, jump-diffusive noise sources do have a wide range of applications. The number of potential applications is naturally multiplied if we consider $\lambda$ and/or $\sigma$ as space dependent. Space correlations in the noise sources typically occur in the mean-field description of interacting particle systems and multi-agents modeling. We have in mind applications, where simple mutual interactions between agents (resp. particles) give rise to mean-field dynamics for the barycenter of the spatially distributed agents. Recent contributions relevant for our context here are the results derived  by  M. Bal\'azs et al. \cite{BALAZS} and the applications in  \cite{HONGFILGAL} and \cite{HONGLER15}. These papers show that, under adequate conditions, the stationary barycentric dynamics of multi-agents systems develop traveling wave solutions.
Generalizing on these results, we consider the case $f(x)=0$, and $\sigma=0$ for $m=1$ and $m=2$. According to Eq.(\ref{DIFFOSALT2}), one immediately  has:
\begin{eqnarray}
\label{MULTIA}
\left[ \partial_x + \gamma \right] \big( \partial_t P_1  \big) &=&
- \partial_x   \big(\lambda \cdot P_1 \big),\;\;(m=1)\\
\left[ \partial_x + \gamma \right]^{2} \big( \partial_t P_2 \big) &=&
-\partial_{x}\big[\partial_{x} + 2\gamma  \big]  \big(\lambda \cdot P_2 \big),\;(m=2)
\end{eqnarray}

\noindent In the sequel, we shall assume that the shot noise rate $\lambda(x,t)$ is a strictly positive and monotone decreasing function in $x$, thereby potentially giving rise to traveling wave-type stationary distributions. For such a stationary propagating regime we will have  $\lim_{t \rightarrow \infty} \mathbb{E}_m\left\{ X(t)\right\} = C_mt$, where $C_m$ is a constant velocity and where $\mathbb{E}_m\left\{ X(t)\right\} := \int_{\mathbb{R}} x P_m dx$. We therefore introduce the change of variable $\xi = x - C_mt$ and suppose the for large $t$, the jump rate is of the form:
\begin{equation}
\label{JUMP}
\lambda(x,t) = \lambda(x - \mathbb{E}_m\left\{ X(t)\right\} )=\lambda(\xi) \geq 0.
\end{equation}
\noindent Under these assumptions, the equations in (\ref{MULTIA}) can be rewritten as ODE's in $\xi\in \mathbb{R}$.

\vspace{0.3cm}
\noindent  $\bullet$ For $m=1$, we have:
 \begin{equation}
\label{STATIOE1}
- C_1(\gamma + \partial_{\xi}) \partial_{\xi} P_1(\xi) = -\partial_{\xi} \left\{  \left[  \lambda (\xi) P_1(\xi)\right] \right\}\;, \;\;(m=1)
\end{equation}
\noindent  admitting the traveling wave solution $P_1(\xi) = {\cal N} e^{- \gamma \xi + \int^{\xi} { \lambda (z) dz \over C_1} }$ with $\cal N$ being the normalization constant which must be self-consistently determined under the constraint $\int_{\mathbb{R}} \xi\cdot  P_1(\xi) d\xi=0$.

\vspace{0.3cm}
\noindent  $\bullet$ For $m=2$, we have after one immediate integration with respect to $\xi$:
 \begin{equation}
\label{STATIOE}
- C_2(\gamma + \partial_{\xi})^2  P_2(\xi) = - \left\{ 2\gamma + \partial_{\xi} \right\}\left[  \lambda (\xi) P_2(\xi)\right] ,\;\;(m=2)
\end{equation}
\noindent  which, if we introduce the auxiliary function $\Psi(\xi)$ defined through:
\begin{equation}
\label{NEWVARI}
P_2(\xi) = \exp\left\{ - \gamma \xi + \int^{\xi} {\lambda (z)  \over 2C_2} dz \right\} \Psi(\xi),
\end{equation}
\noindent reduces to
\begin{equation}
\label{PSI}
\partial_{\xi \xi} \Psi(\xi) + \left[-{\partial_{\xi} \lambda(\xi) \over 2 C_2} - { \lambda^{2}(\xi)\over 4 C_{2}^{2}} -{\gamma \lambda(\xi) \over C_2}\right] \Psi(\xi)=0.
\end{equation}

\noindent  We observe that for arbitrary $\lambda(\xi)$, Eq.(\ref{PSI}) exhibits  the form of  a stationary Schr\"odinger equation which, in general, cannot be solved in compact form. Looking for compact solutions to Eq.(\ref{PSI}), the term in brackets can be related to analytically tractable potentials in quantum mechanics. To carry on the discussion for $m=1$ and $m=2$, we focus on the special case which results, when the jump rates are of the form $\lambda(\xi) = e^{- \beta \xi}$, with $\beta>0$.


\subsubsection*{Jump rate  governed by $\lambda(\xi) = e^{- \beta \xi}$.}

\noindent  $\bullet$\quad  For $m=1$,  this case has been worked out  in the mean-field context of an interacting particle systems by Balazs et al. in \cite{BALAZS}, (see the Corollary 3.2),  we find that  $P_1(\xi)$ is a Gumbel-type distribution:
\begin{equation}
\label{GUMBOS}
P_1(\xi)  =  {\cal N}(\beta, \gamma, C_1)e^{ - \gamma \xi - {1 \over \beta C_1}e^{- \beta \xi} },
\end{equation}
\noindent   with $ {\cal N}(\beta, \gamma, C_1)$ being the normalization factor. The normalization $ {\cal N}$ and the resulting stationary velocity $C_1$ are explicitly found to be:
\begin{eqnarray}
\label{C}
{\cal N}(\beta, \gamma, C_1)  &=& \frac{\beta}{(\beta C_1)^{\frac{\gamma}{\beta}}\Gamma(\gamma/\beta)}\\
C _1 &=& {1 \over \beta } e^{-\psi (\gamma/\beta)}, \textrm{with}\;\;  \psi(x) := {d\over dx} \ln \left[\Gamma(x) \right]
\end{eqnarray}
ensuring $\int_{\mathbb{R}} P_1(\xi)  d\xi= 1$ and $\int_{\mathbb{R}} \xi P_1(\xi)  d\xi= 0$.

\vspace{1cm}
\noindent $\bullet$\quad For $m=2$,  Eq.(\ref{PSI}) now reads:

\begin{equation}
\label{GUMBEL}
\partial_{\xi \xi} \Psi(\xi) + \left[{(\beta - 2 \gamma) \over 2 C_2} e^{- \beta \xi}- {1\over 4 C_{2}^{2}} e^{- 2\beta \xi}\right] \Psi(\xi)=0.
\end{equation}

\noindent Observe that Eq.(\ref{GUMBEL}) corresponds to the stationary Schr\"odinger Eq. describing a quantum particle submitted to a Morse type potential for which explicit solutions are known. Using these results in the expression for $P_2(\xi)$  and imposing vanishing boundary conditions for large $|\xi|$ (see Appendix B for details), we find:
\begin{equation}
\label{SOLVAR}
\begin{array}{l}
P_2(\xi) ={\cal N}(\beta, \gamma, C_2)e^{\left[ { \beta\over 2} - \gamma \right]\xi - {e^{- \beta \xi }\over 2 \beta C_2}}  W_{{\beta - 2 \gamma \over 2\beta }, 0}\left({e^{- \beta \xi}\over \beta C_2}\right)
\end{array}
\end{equation}
where $W_{\lambda, \mu}(z)$ is the Whittaker $W$ function (see \cite{GRADSHTEYN} 9.22) and $ {\cal N}(\beta, \gamma, C_2)$ is the normalization factor. The normalization $ {\cal N}$ and the resulting stationary velocity $C_2$ are explicitly found to be:
\begin{eqnarray}
\label{C}
{\cal N}(\beta, \gamma, C_2)  &=& \frac{\beta}{(\beta C_2)^{\frac{\gamma}{\beta}-\frac{1}{2}}}\frac{\Gamma(2\gamma/\beta)}{\Gamma(\gamma/\beta)^2}\\
C _2 &=& {1 \over \beta } e^{\psi (2\gamma/\beta)-2\psi(\gamma/\beta)},
\end{eqnarray}
ensuring $\int_{\mathbb{R}} P_2(\xi)  d\xi= 1$ and $\int_{\mathbb{R}} \xi P_2(\xi)  d\xi= 0$. It is worthwhile noting that $C_2/C_1=exp(e^{\psi (2\gamma/\beta)-\psi(\gamma/\beta)})>2$ showing explicitly how the jump size parameter $m$ influences the speed of the traveling wave solution $P_m(\xi)$.

\begin{figure}[h]\vspace*{0mm}
  \centering
  \includegraphics[width=12cm,height=5cm]{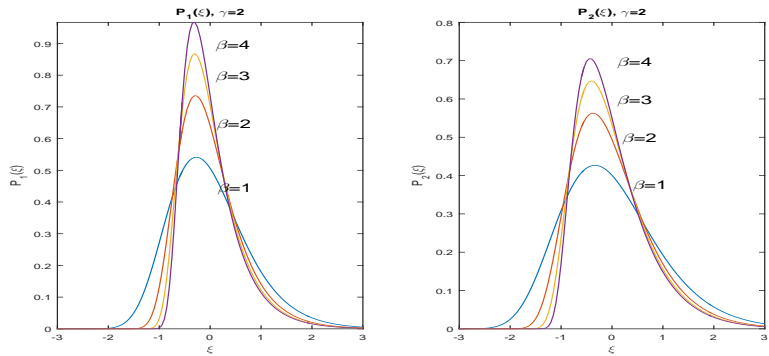}
  \caption{Exact normalized traveling probability waves $P_1(\xi) $ and $P_2(\xi) $ as given by Eqs.(\ref{GUMBOS}) respectively (\ref{SOLVAR}) for different values of $\beta$. }
\end{figure}

\section{Exactly soluble nonlinear mixed jump-diffusive  processes}

The mixed jump-diffusive processes defined by \eqref{SDE1} do have the Markov property and are, under the assumption of sufficient symmetries, lumpable to simpler processes \cite{PITMAN}. In the realm of lumbaple Markov diffusions, an outstanding role is played by Brownian motions with drift of the form $f(x)=\beta \tanh(\beta x)$ as they are, together with the class of Brownian motions with constant drift, the only ones having Brownian bridges as conditional laws \cite{BENJAMINI}. This non linear and lumpable drift offers indeed the exceptional possibility to escape in a controlled and still analytical way from the Gaussian law (see e.g., \cite{HONGLER81, HONGLER06, HONGLER08}). We therefore consider the 1 dimensional dynamics given by:

\begin{equation}
\label{NLSN2}
\left\{
\begin{array}{l}
dX_t=  \beta \tanh[\beta X_t] dt+   dW_t + q_{t}, \\ \\
X_0 =x_0,
\end{array}
\right.
\end{equation}
\noindent where in this section $q_{t}$ is a Poisson process with constant rate $\lambda$ and jump sizes drawn from a symmetric probability law $ \phi(x)$ (i.e., respecting $\phi(x) = \phi(-x)$ and $\int_{-\infty}^{\infty} \phi(x)  dx =1$). We therefore can have positive and negative jumps. The Master equation related to Eq.(\ref{NLSN2}) reads:
\begin{equation}
\label{KOLON}
\begin{array}{l}
{\partial \over \partial t} Q(x,t|x_0)  = - \beta {\partial \over \partial x} \left\{  \tanh(\beta x) Q(x,t|x_0) \right\} + {1 \over 2} \partial_{xx} Q(x,t|x_0) \\
\hspace*{20mm} - \lambda  Q(x,t|x_0) + \lambda \int_{-\infty}^{x} Q(x-y, t|x_0)  \phi(y)dy.
\end{array}
\end{equation}
\noindent By introducing the transformation  $Q(x,t|x_0) =e^{-{1 \over 2} \beta^{2} t} \cosh (\beta x) R(x,t|x_0)$,   it is immediate to verify that Eq.(\ref{KOLON}) takes the form:
\begin{equation}
\label{REMOKOLON}
\begin{array}{l}
{\partial \over \partial t} R(x,t|x_0)  =   {1 \over 2} \partial_{xx} R(x,t|x_0) -\lambda   R(x,t|x_0)  \\ \\
\qquad  \quad +  {\lambda \over \cosh(\beta x)} \int_{-\infty}^{x} \cosh\left[\beta (x -y)\right]  R(x-y, t|x_0)  \phi(y) dy.
\end{array}
\end{equation}
\noindent The identity $\cosh(a + b) = \cosh(a) \cosh (b) + \sinh(a) \sinh (b),$ enables to rewrite   Eq.(\ref{REMOKOLON})   as:
\begin{equation}
\label{REMOKOLON1}
\begin{array}{l}
{\partial \over \partial t} R(x,t|x_0)  =  {1 \over 2} \partial_{xx} R(x,t|x_0)
 + \lambda   \int_{-\infty}^{x} R(x-y, t|x_0)  \phi(y) \cosh(\beta y) dy \\ \\
\qquad  - \lambda  R(x,t|x_0) - {1 \over 2}\lambda \tanh(\beta x) \int_{-\infty}^{x}  \sinh(\beta y)  R(x-y, t|x_0)  \phi(y) dy.
\end{array}
\end{equation}
\noindent  When the initial condition  is taken $x_0=0$, symmetry of $\phi$ implies
$Q(x,t|0) = Q(-x,t|0)$ and therefore also $R(x,t|0) = R(-x,t|0)$. Accordingly, when $x_0=0$, the second integral in  Eq.(\ref{REMOKOLON1}) vanishes and hence Eq.(\ref{REMOKOLON1})  describes the evolution of the TPD $R(x,t|0)$ which characterizes a drift-free jump diffusion process  $\tilde{X}(t)$, solution of  %
\begin{equation}
\label{NLSN}
{d \over dt} \tilde{X}(t) =  dW_t + q_{\beta,t}
\end{equation}
\noindent where now the  Poisson noise $ q_{\beta,t}$ is characterized by  jumps drawn from the probability law $  \phi_{\beta} (x) := \phi(x) \cosh(\beta x)$. Let us write $Q_{\beta} (x,t|0)$ for the TPD associated with the jump part in Eq.(\ref{NLSN}). Then we can write:

\begin{equation}
\label{QBETA}
R(x,t|0) = {\cal N}(x,t|0) \ast Q_{\beta}(x,t|0)
\end{equation}

\noindent where $\ast$ stands for the convolution and  where ${\cal N}(x,t) :=( \sqrt{2 \pi t})^{-1} e^{- {x^{2} \over 2t}}$. Finally, for $x_0=0$, the TPD $Q (x,t|0)$ solving Eq.(\ref{KOLON}) reads:

\begin{equation}
\label{CON}
\left\{
\begin{array}{l}
Q(x,t|0) = e^{-{1 \over 2} \beta^{2} t} \cosh (\beta x) {\cal N}(x,t|0) \ast Q_{\beta}(x,t|0)\\ \\Ê
\qquad \qquad \qquad \qquad ={1 \over 2} \left[ {\cal N}^{(+\beta)} (x,t|0)+ {\cal N}^{(-\beta)}(x,t|0)\right] \ast Q_{\beta}(x,t|0),
\end{array}
\right.
\end{equation}

\noindent with ${\cal N}^{(\pm \beta)}(x,t) :=( \sqrt{2 \pi t})^{-1} e^{- {(x- \beta t)^{2} \over 2t}}$.
\vspace{1cm}
%

\noindent {\bf Ilustration}. The superposition of probability measures given by Eq.(\ref{CON}) can be used to derive explicitly new probability measures. For example,  let us consider the case where in Eq.(\ref{KOLON}) we take:

\begin{equation}
\label{BISIDE}
\phi(x ) = {\gamma \over 2} e^{- \gamma |x|}.
\end{equation}

\noindent and for  this  choice, we consider the generalized Ornstein-Uhlenbeck dynamics $Y_t$ characterized by:

\begin{equation}
\label{NEWLINDYN}
dY_t = - \alpha Y_t dt + dX_t ,
\end{equation}

\noindent where in Eq.(\ref{NEWLINDYN})  the noise source $dX_t$ is given by Eq.(\ref{NLSN2}). The superposition given in Eq.(\ref{CON}) enables to write the TPD $P(y,t|y_0)$ characterizing  the process $Y_t$ as:

\begin{equation}
\label{LUMPOS}
 P(y,t|y_0) = {1 \over 2}  \left[ P^{(+\beta)}(y,t|y_0)  + P^{(-\beta)}(y,t|y_0) \right],
\end{equation}
\noindent where $ P^{(\pm\beta)}(y,t|y_0)$ are the TPD of the respective  processes:
\begin{equation}
\label{LUMPOSPM}
\left\{
\begin{array}{l}
dY^{(\beta)}_t = - \alpha Y_t dt + dX_{\beta, t}\\ \\
dX_{\beta, t} =  \pm \beta dt +   dW_t + q_{t}
\end{array}
\right.
 \end{equation}

 \noindent where $q_{t}$ is the pure jump process with Poisson rate $\lambda$ and jump size distribution ${1 \over 2}\gamma e^{- \gamma |x|}$. Using the results derived in \cite{DALEY06, DENISOV}, we have \footnote{See for instance Eq. (13) in \cite{DALEY06}.}:
 \begin{equation}
\label{MESINV}
\left\{
\begin{array}{l}
\lim_{t \rightarrow \infty} P^{(\pm \beta)}(y,t|y_0):= P^{(\pm \beta)}_{s}(y) = {2^{\nu} \gamma ^{1- \nu}  |y \pm {\beta\over \alpha}|^{- \nu} \over \sqrt{\pi} \, \Gamma\left[{1\over 2}- \nu \right]} \mathbb{K}_{\nu}(\gamma |y \pm {\beta\over \alpha}|),\\ \\
\nu:={1 \over 2} \left[ 1 - {\lambda \over \alpha }\right],
\end{array}
\right.
\end{equation}
where $\mathbb{K}_{\nu}$ is the modified Bessel function of the second kind.
\noindent Consequently, the invariant measure $P_{s}(y)$ for the process Eq.(\ref{LUMPOSPM}) reads as:

\begin{equation}
\label{RESULT}
\lim_{t \rightarrow \infty} P(y,t|y_0) = P_{s}(y) = {1 \over 2} \left[ P^{(- \beta)}_{s}(y) + P^{(+ \beta)}_{s}(y)\right] .
\end{equation}

\section*{Conclusion}

Jump diffusions offer a rich class of noise sources and are widely used as modeling tools in various fields. As such, special interest lies in the explicit understanding of the effect of different jump distributions
on the model dynamics. It is remarkable that in cases of space inhomogeneous shot noise with jump sizes following a gamma distribution with parameter $(m,\gamma)$, and space inhomogeneous jump frequency, $\lambda(\xi)=e^{- \beta \xi}$, a differential form of the Master-equation allows to quantitatively unveil the influence of the shape parameter $m$ on the speed of stationary traveling wave solutions.


\section*{Appendix A}
\noindent To the readers convenience, we give a detailed proof of proposition 1. We proceed by induction over $m\in \mathbb{N}$ (the Erlang parameter). We indeed show that \eqref{DIFFOSALT2} follows from  \eqref{ERLANGm} by applying the operator
$\mathcal{O}_m:=e^{-\gamma x}\partial_{x}^m e^{\gamma x}(\cdot )$ to \eqref{ERLANGm}, where $\partial_{x}^m$ is the $m$-fold derivative with respect   to $x$.

\noindent We start with the basic case by direct calculation and apply $\mathcal{O}_m$ to \eqref{ERLANGm} for $m=1$ and use, for notational ease, $f(\cdot)=f$, $\lambda(\cdot,\cdot)=\lambda$, $\sigma(\cdot,\cdot)=\sigma$ and likewise $\partial_{x}(\cdot)$ or $(\cdot)_x$ for derivatives wrt $x$. We find:
{\scriptsize
\begin{eqnarray}
\label{A:ER1}
e^{-\gamma x}\partial_{x} e^{\gamma x}\Big(\partial_t P_1-(f P_1)_x-(\frac{\sigma^2}{2} P_1)_{x,x}\Big)&= &e^{-\gamma x}\partial_{x} e^{\gamma x}\Big( - \lambda P_1 + \gamma \int_{0}^{x} e^{- \gamma(x-z)} \lambda P_1(z)dz\Big) \nonumber \\
\left[ \gamma + \partial_x \right] \big(\partial_t P_1-(f P_1)_x-(\frac{\sigma^2}{2} P_1)_{x,x}\big)  &=& e^{-\gamma x}\Big( - \big(\gamma e^{\gamma x} \lambda P_1 +  e^{\gamma x} (\lambda P_1)_{x}\big)+\gamma e^{\gamma x} \lambda P_1\Big) \nonumber \\
\left[ \gamma + \partial_x \right] \big(\partial_t P_1-(f P_1)_x-(\frac{\sigma^2}{2} P_1)_{x,x}\big)  &=&  -    (\lambda P_1)_{x} \nonumber
\end{eqnarray}}

\noindent which matches the proposition for $m=1$.

\noindent For the induction step, we note $I_m$ for the integral part of \eqref{ERLANGm}, i.e.:
$$I_m=\int_{0}^{x} {\gamma ^{m} (x-z)^{m-1} e^{- \gamma (x-z)}\over \Gamma (m)} \lambda(z,t) P_m(z, t |x_0,0) dz$$
and remark that $\partial_{\gamma} I_m=\frac{m}{\gamma}I_m-\frac{m}{\gamma}I_{m+1}.$ Hence,
\begin{eqnarray}\label{Im}I_{m+1}=I_m-\frac{\gamma}{m}\partial_{\gamma}I_m\end{eqnarray}
Let us apply $\mathcal{O}_{m+1}=e^{-\gamma x}\partial_{x}^{m+1} e^{\gamma x}(\cdot )$ to \eqref{ERLANGm} for the case $m+1$. Using the Leibnitz formula for higher order derivatives of productes\footnote{The Leibnitz formula $\partial_x^m(f(x)g(x))=\sum_{k=0}^m\binom{m}{k}f^{(k)}g^{(m-k)}$, with $f(x)=e^{\gamma x}$ takes the form $\partial_x^m(e^{\gamma x}g(x))=e^{\gamma x}\sum_{k=0}^m\binom{m}{k}\gamma^{k}g^{(m-k)}$ and with the binomial formula we get $\partial_x^m(e^{\gamma x}g(x))=e^{\gamma x}(\partial_x+\gamma)^{m}g(x)$}, the left hand side is immediately seen to be
$$\left[ \gamma + \partial_x \right]^{m+1} \big(\partial_t P_{m+1}-(f P_{m+1})_x-(\frac{\sigma^2}{2} P_1)_{x,x}\big).$$
Apply the operator $\mathcal{O}_{m+1}$ to the right hand side of \eqref{ERLANGm} and use \eqref{Im} to establish:
{\small\begin{eqnarray}\label{I3}
&&\hspace*{-10mm}e^{-\gamma x}\partial_{x}^{m+1} e^{\gamma x} \Big( - \lambda P_{m+1}(x,t)  + I_{m+1}\Big) =\nonumber \\
&&\hspace*{10mm} e^{-\gamma x}\partial_{x}e^{\gamma x}\Big\{e^{-\gamma x}\partial_{x}^{m} e^{\gamma x} \Big\}\Big( - \lambda P_{m+1}(x,t)  + I_{m}- \frac{\gamma}{m}\partial_{\gamma}I_m\Big)\nonumber
\end{eqnarray}}
Within the brackets we recognize $\mathcal{O}_m$ which acts upon the left hand side of \eqref{ERLANGm} for $m$ and also on the extra term ($-\frac{\gamma}{m}I_m$). Using the induction hypothesis, the right hand side reads:
\begin{eqnarray}\label{I4}
e^{-\gamma x}\partial_{x}e^{\gamma x}\Big(\big[\gamma^{m}-\left[ \partial_x + \gamma \right]^{m}\big]  \big(\lambda \cdot P_{m+1} \big)- e^{-\gamma x}\partial_{x}^{m} e^{\gamma x} \frac{\gamma}{m}\partial_{\gamma}I_m\Big)\nonumber
\end{eqnarray}
A direct computation of the last term in the above parenthesis gives:

$$e^{-\gamma x}\partial_{x}^{m} e^{\gamma x} \frac{\gamma}{m}\partial_{\gamma}I_m=\gamma^{m}\lambda P_{m+1}-\int_0^x\gamma^{m+1}e^{-\gamma(x-z)}\lambda P_{m+1}dz.$$
We therefore are left to show hat
{\scriptsize
\begin{eqnarray}\label{I5}
\hspace*{-10mm}e^{-\gamma x}\partial_{x}e^{\gamma x}\Big(\big[\gamma^{m}-\left[ \partial_x + \gamma \right]^{m}\big]  \big(\lambda \cdot P_{m+1} \big)-\gamma^{m}\lambda P_{m+1}
+ \int_0^x\gamma^{m+1}e^{-\gamma(x-z)}\lambda P_{m+1}dz.\Big)\nonumber\\
\hspace*{20mm}\stackrel{!}{=}\big[\gamma^{m+1}-\left[ \partial_x + \gamma \right]^{m+1}\big]  \big(\lambda \cdot P_{m+1} \big)
\end{eqnarray}}

\noindent For the first term we get:
{\scriptsize
\begin{eqnarray}\label{I6}
\hspace*{-10mm}e^{-\gamma x}\partial_{x}e^{\gamma x}\big[\gamma^{m}-\left[ \partial_x + \gamma \right]^{m}\big]  \big(\lambda \cdot P_{m+1} \big) &=& \gamma^{m}\left[ \partial_x + \gamma \right]\big(\lambda \cdot P_{m+1} \big)- \left[ \partial_x + \gamma \right]^{m+1}\big(\lambda \cdot P_{m+1} \big) \nonumber\\
&&\hspace*{-35mm}= \big[\gamma^{m+1}-\left[ \partial_x + \gamma \right]^{m+1}\big]  \big(\lambda \cdot P_{m+1} \big)+\partial_x \big(\gamma^m \lambda P_{m+1}\big)\nonumber
\end{eqnarray}}
For the middle term we find:

\begin{eqnarray}\label{I7}
e^{-\gamma x}\partial_{x}e^{\gamma x}\Big(-\gamma^{m}\lambda P_{m+1}\Big)=-\gamma^{m+1}\lambda P_{m+1}-\partial_x \big(\gamma^m \lambda P_{m+1}\big)
\end{eqnarray}
finally the last term is:

\begin{eqnarray}\label{I8}
e^{-\gamma x}\partial_{x}e^{\gamma x}\Big(\int_0^x\gamma^{m+1}e^{-\gamma(x-z)}\lambda P_{m+1}dz\Big)=\gamma^{m+1}\lambda P_{m+1}
\end{eqnarray}
Hence, adding \eqref{I6}-\eqref{I8} together, we have established \eqref{I5} and therefore also proposition 1.


\section*{Appendix B}
\noindent Our starting point is Eq.(\ref{GUMBEL}), (issued from Eq.(\ref{PSI}) when $\lambda(\xi) = e^{- \beta \xi}$).

\noindent First we introduce the change of variable

\begin{equation}
\label{CVAR}
\left\{
\begin{array}{l}
Z = e^{- \beta \xi} \quad \Rightarrow \quad dZ = - \beta Z d\xi, \\ \\
\partial_{\xi} (\cdot)  \mapsto - \beta Z \partial_{Z}(\cdot), \\ \\
\partial_{\xi  \xi } (\cdot) \mapsto  \beta^{2} Z^{2} \partial_{Z Z}(\cdot)  + \beta^{2} \partial_Z(\cdot) .
\end{array}
\right.
\end{equation}

\noindent In terms of the $Z$-variable, Eq.(26) takes the form:

\begin{equation}
\label{WHIT1}
\left\{
\begin{array}{l}
\beta^{2} Z^{2} \partial_{Z Z}\Psi(Z)  + \beta^{2} \partial_Z\Psi(Z) + \left[ q Z - pZ^{2}\right] \Psi(Z)=0, \\ \\
q:= {(\beta - 2\gamma)\over 2 C_2}\qquad {\rm and}\qquad  p:= {1 \over 4C_2^{2}}.
\end{array}
\right.
\end{equation}

\noindent Or equivalently:

\begin{equation}
\label{WHIT2}
 \partial_{Z Z}\Psi(Z)  + {1 \over Z} \partial_Z\Psi(Z) + \left[ {q \over \beta^{2} Z} - {p \over \beta^{2}}\right] \Psi(Z)=0. \\ \\
\end{equation}

\noindent Let us now write:

\begin{equation}
\label{NEWCVAR}
\Psi(Z) = Z^{-{1 \over 2}} \varphi(Z).
\end{equation}

\noindent Accordingly $ \varphi(Z)$ obeys to the equation:

\begin{equation}
\label{MODWHIT1}
 \partial_{Z Z}\varphi(Z)  + \left[{1 \over 4 Z^{2} }+ {q \over \beta^{2} Z} - {p \over \beta^{2}} \right] \varphi(Z)=0.
\end{equation}

\noindent Now, let us introduce the rescaling:

\begin{equation}
\label{RESCALE}
\left\{
\begin{array}{l}
U = \omega Z, \\ \\
 \partial_{Z}(\cdot) \mapsto \omega \partial_{U} (\cdot)\qquad {\rm and} \qquad \partial_{ZZ}(\cdot) \mapsto \omega^{2} \partial_{UU}(\cdot)
\end{array}
\right.
\end{equation}

\noindent Using Eq.(\ref{RESCALE}) in Eq.(\ref{MODWHIT1}), we obtain:

\begin{equation}
\label{STANDARWHIT}
\partial_{UU}\varphi(U)  + \left[{1 \over 4 U^{2} }+ {q \over \omega \beta^{2} U} - {p \over  \omega^{2}\beta^{2}} \right] \varphi(U)=0
\end{equation}

\noindent Now, to match the standard Whittaker equation, (see entry 13.1.13 of \cite{ABRAMOWITZ}), we have to select:

\begin{equation}
\label{SCALE}
{p \over  \omega^{2}\beta^{2}}  = {1 \over 4} \quad \Rightarrow \quad \omega = {1 \over \beta C_2}.
\end{equation}

\noindent So the general  solution of Eq.(\ref{WHIT1}) reads:

\begin{equation}
\label{SOL}
\Psi(\xi) = \sqrt{\beta C_2} \, {e^{\beta \xi \over 2}} \left\{ A\,  M_{{\beta - 2 \gamma \over2 \beta },0} \left({e^{- \beta \xi}\over \beta C_2}\right) + B\, W_{{\beta - 2 \gamma \over 2\beta }, 0}\left({e^{- \beta \xi}\over \beta C_2}\right)\right\},
\end{equation}

\noindent where $A$ and $B$ are yet undetermined constants. By using Eq.(21), the probability density $P_2(\xi)$  reads:

\begin{equation}
\label{PDEUX}
\begin{array}{l}
P_2(\xi) = {\cal N} e^{- \gamma \xi - {e^{- \beta \xi }\over 2 \beta C_2}} \Psi(\xi) = \\ \\
\qquad \qquad  {\cal N} e^{\left[ { \beta\over 2} - \gamma \right]\xi - {e^{- \beta \xi }\over 2 \beta C_2}} \left\{ A\,  M_{{\beta - 2 \gamma \over2 \beta },0} \left({e^{- \beta \xi}\over \beta C_2}\right) + B\, W_{{\beta - 2 \gamma \over 2\beta }, 0}\left({e^{- \beta \xi}\over \beta C_2}\right)\right\}.
\end{array}
\end{equation}

\noindent   where ${\cal N}$ is the normalization factor. Let us now calculate the  average of the positive definite function ${\cal G}(u)$ defined as:

\begin{equation}
\label{GENEROUS}
{\cal G}(u) := \int_{-\infty}^{+\infty} e^{- u\xi}P_2(\xi) d\xi >0.
\end{equation}

\noindent  and the normalization imposes that  ${\cal G}(0)=1$.
\noindent Now, we introduce the new  variable ${\cal Z}$ defined as:

\begin{equation}
\label{CALZ}
{\cal Z} := {e^{-\beta \xi} \over \beta },
\end{equation}

\noindent  In terms of this new variable, Eq.(\ref{PDEUX}) now reads:

\begin{equation}
\label{NNN}
\begin{array}{l}
{\cal G}(u) =   {\cal N}\int_{0}^{\infty} e^{{ {\cal Z}\over 2  C_2}} {\cal Z}^{{\gamma +u \over  \beta}- {3 \over 2}} \left\{ A \, M_{{1 \over 2}  -  {\gamma \over \beta}, 0} \left({{\cal Z} \over C_2}\right) +
B\, W_{{1 \over 2}  -  {\gamma \over \beta}, 0}\left({{\cal Z} \over C_2}\right) \right\} d{\cal Z}.\end{array}
\end{equation}

\noindent Now we use, the entries  7.622.8 and 7.622.11 from I. S. Gradshteyn to calculate ${\cal I}_1$ and  ${\cal I}_e$ with the choice of parameters $b= {1 \over C_2}$, $\mu =0$, $\nu = {\gamma +u\over \beta} - {1 \over 2} $ and $\kappa = {1 \over 2}  -  {\gamma \over \beta}$ leading to :

\begin{equation}
\label{GENEROUS1}
\begin{array}{l}
{\cal G}(u)= A \left[ {\Gamma \left( 1- {2 \gamma + u \over \beta }\right) \Gamma \left( { \gamma + u \over \beta }\right)  \Gamma \left( 1- { \gamma \over \beta }\right) \over   \Gamma \left( 1- {2 \gamma \over \beta }\right) \Gamma \left( { \gamma  \over \beta }\right)\Gamma \left( 1- { \gamma + u \over \beta }\right) }\right] (C_2)^{-{u \over \beta}} + \\ \\ 
\qquad \qquad \qquad \qquad\qquad \qquad \quad  
B  \left[
 {  \Gamma \left( {2 \gamma \over \beta }\right)  
 \Gamma \left( { \gamma + u \over \beta }\right)^{2}  
 \over  
 \Gamma \left( { \gamma  \over \beta }\right)^{2}  \Gamma \left( {2 \gamma + u \over \beta }\right) 
  }\right] (C_2)^{{u \over \beta}}.
\end{array}\end{equation}

\noindent As ${\cal G}(u) >0$, the arguments of the Gamma functions have to be strictly positive for all values of $\gamma$ and $\beta$. Hence, we are forced to impose  $A=0$ and hence $B=1$.  Let us now calculate the velocity $C_2$, we end with

\begin{equation}
\label{ AVERAGEC}
\begin{array}{l}
C_2 =0= - {d \over du} {\cal G}(u)\mid_{u=0} =  - {d \over du} \left[ e^{{u \over \beta}\ln(C_2)}  \varphi(u)\right]\mid_{u=0} \quad \Rightarrow \\ \\
\quad  \qquad \qquad \qquad C_2 ={1\over \beta }e^{{1 \over \beta} \left[\Psi\left({2 \gamma\over \beta}\right) - 2 \Psi\left({\gamma \over \beta}\right)\right]},
\end{array}
\end{equation}

\noindent where $\Psi(x) : = {d \over dx} \ln [\Gamma(x) ]$ is the digamma function

\bibliographystyle{unsrt}
\bibliography{SOLOW}

\begin{thebibliography}{10}

\bibitem{COX}
D.~R. Cox and H.~D. Miller.
\newblock The theory of stochastic processes.
\newblock 1965.

\bibitem{DENISOV}
S.~I. Denisov, H.~Kantz, and P.~H{\"a}nggi.
\newblock Langevin equation with super-heavy-tailed noise.
\newblock {\em Journal of Physics A: Mathematical and Theoretical}, 43(28),
  2010.

\bibitem{Denisov2009}
I.~S. Denisov, W.~Horsthemke, and P.~H{\"a}nggi.
\newblock Generalized fokker-planck equation: Derivation and exact solutions.
\newblock {\em The European Physical Journal B}, 68(4):567--575, 2009.

\bibitem{ELIAZAR}
I.~Eliazar and J.~Klafter.
\newblock On the nonlinear modeling of shot noise.
\newblock {\em Proceedings of the National Academy of Sciences of the United
  States of America}, 102(39):13779--13782, 2005.

\bibitem{DALEY}
E.~Daly and A.~Porporato.
\newblock Effect of different jump distributions on the dynamics of jump
  processes.
\newblock {\em Physical Review E - Statistical, Nonlinear, and Soft Matter
  Physics}, 81(6), 2010.

\bibitem{DALEY06}
E.~Daly and A.~Porporato.
\newblock Probabilistic dynamics of some jump-diffusion systems.
\newblock {\em Physical Review E - Statistical, Nonlinear, and Soft Matter
  Physics}, 73(2), 2006.

\bibitem{BALAZS}
M.~Bal\'azs, M.~Z. R\'acz, and B.~T\'{o}th.
\newblock Modeling flocks and prices: Jumping particles with an attractive
  interaction.
\newblock {\em Annales de l'institut Henri Poincare (B) Probability and
  Statistics}, 50(2):425--454, 2014.

\bibitem{PERRY}
D.~Perry, W.~Stadje, and S.~Zacks.
\newblock First-exit times for poisson shot noise.
\newblock {\em Communications in Statistics.Part C: Stochastic Models},
  17(1):25--37, 2001.

\bibitem{ABRAMOWITZ}
M.~Abramowitz and I.~Stegun.
\newblock {\em Handbook of mathematical functions}.
\newblock Dover. 1964.

\bibitem{Takacs}
L.~Takács.
\newblock The transient behavior of a single server queuing process with a
  poisson input.
\newblock {\em Proceedings of the Fourth Berkeley Symposium on Mathematical
  Statistics and Probability}, 2:pp. 535–567, 1961.

\bibitem{HONGFILGAL}
M.-O. Hongler, R.~Filliger, and O.~Gallay.
\newblock Local versus nonlocal barycentric interactions in 1d dynamics.
\newblock {\em Mathematical Bioscience and Engineering}, 11(2):323--351, 2014.

\bibitem{HONGLER15}
M.-O. Hongler.
\newblock Exact soliton-like probability measures for interacting jump
  processes.
\newblock {\em Mathematical Scientist}, 40(1):62--66, 2015.

\bibitem{GRADSHTEYN}
I.~S. Gradshteyn and M.~Ryzhik.
\newblock {\em Tables of integrals, series and products}.
\newblock Academic Press. 1980.

\bibitem{PITMAN}
J.~Pitman and L.C.G. Rogers.
\newblock Markov functions.
\newblock {\em The Annals of Probab.}, 9(4):573,582, 1981.

\bibitem{BENJAMINI}
I.~Benjamini and S.~Lee.
\newblock Conditioned diffusions which are brownian bridges.
\newblock {\em Journal of Theoretical Probability}, 10(3):733--736, 1997.

\bibitem{HONGLER81}
M.-O. Hongler.
\newblock Study of a class of non-linear stochastic processes boomerang
  behaviour of the mean path.
\newblock {\em Physica D: Nonlinear Phenomena}, 2(2):353--369, 1981.

\bibitem{HONGLER06}
M.-O. Hongler, R.~Filliger, and P.~Blanchard.
\newblock Soluble models for dynamics driven by a super-diffusive noise.
\newblock {\em Physica A: Statistical Mechanics and its Applications},
  370(2):301--315, 2006.

\bibitem{HONGLER08}
M.-O. Hongler and P.~R. Parthasarathy.
\newblock On a super-diffusive, nonlinear birth and death process.
\newblock {\em Physics Letters, Section A: General, Atomic and Solid State
  Physics}, 372(19):3360--3362, 2008.

\end{thebibliography}

\end{document}